\documentclass[aps,showpacs,preprintnumbers,amsmath,amssymb,nofootinbibt,preprint]{article}
\usepackage{graphicx}

\begin{document}

\title{\hfill {\small\bf ECTP-2009-8}\\
Dissipative Processes in the Early Universe: Bulk Viscosity}

\author{A.~Tawfik$^{1}\;$\thanks{drtawfik@mti.edu.eg}$\,$, T.~Harko$^2\,$,H.~Mansour$^3\,$, M.~Wahba$^1\,$ \\~\\
{\small $^1$ Egyptian Center for Theoretical Physics (ECTP), MTI University,
 Cairo-Egypt} \\
{\small $^2$ Department of Physics and Center for Theoretical
and Computational Physics,}\\ {\small University of Hong Kong, Pok Fu Lam Road, Hong Kong}\\
{\small $^3$ Department of Physics, Faculty of Science, Cairo University, Giza-Egypt} \\
}

\date{\today}
\maketitle
\begin{abstract}

In this talk, we discuss one of the dissipative processes which likely take place in the Early Universe. We assume that the matter filling the isotropic and homogeneous background is to be described by a relativistic viscous fluid characterized by an ultra-relativistic equation of state and finite bulk viscosity deduced from recent lattice QCD calculations and heavy-ion collisions experiments. We  concentrate our treatment to bulk viscosity as one of the essential dissipative processes in the rapidly expanding Early Universe and deduce the dependence of the scale factor and Hubble parameter on the comoving time $t$. We find that both scale factor and Hubble parameter are finite at $t=0$, revering to absence of singularity. We also find that their evolution apparently differs from the one resulting in when assuming that the background matter is an ideal and non-viscous fluid.

\end{abstract}

{04.50.Kd, 04.70.Bw, 97.10.Gz}


\section{Introduction}\label{sec:intro}

Dissipative processes are supposed to play a very important role in the Early Universe. The most essential ones are the bulk and shear viscosity. 
First attempts to create a theory of relativistic fluids, which likely simulate the matter under the extreme conditions of the Early Universe, have been done by Eckart \cite{Ec40} and Landau and Lifshitz \cite{LaLi87}. Regardless the choice
of the equations of state, all equilibrium states in these theories (first order) are unstable and in addition signals may propagate through the fluid at velocities
exceeding the speed of light. Therefore these theories are applicable only
to phenomena which are quasi-stationary, i.e. slowly varying on space and
time scales.

A relativistic ''second order'' theory has been suggested by Israel \cite{Is76} and
developed by Israel and Stewart \cite{IsSt76}, Hiscock and Lindblom \cite%
{HiLi89} and Hiscock and Salmonson \cite{HiSa91} into what is called
``transient'' irreversible thermodynamics. In this model
deviations from equilibrium (bulk stress, heat flow and shear stress) are
treated as independent dynamical variables, leading to a total of unknown fourteen 
dynamical fluid variables~\cite{Ma95, Ma96}.

Causal bulk viscous thermodynamics has been used to describe
the dynamics and evolution of the Early Universe. 
But due to the complicated character of the evolution equations,
very few solutions of gravitational field equations
are known in the framework of full causal theory. For a homogeneous
Universe filled with a full causal viscous fluid obeying the relation
$\xi \sim \rho ^{1/2}$~\cite{conseq2}, with $\rho $ being the energy density, 
exact general solutions of the field equations have been obtained in 
\cite{ChJa97}-\cite{MaHa00b}. 
Because of serious technical reasons, most investigations for the dissipative causal
cosmologies have assumed Friedmann-Robertson-Walker (FRW) symmetry or - in few cases - small perturbations 
around it \cite{MaTr97}. 

Recent RHIC results give a strong indication that in the heavy-ion collisions
experiments, a hot dense matter can be formed~\cite{reff1}, which likely agree with the "new state of matter"
as predicted in the Lattice QCD
simulations~\cite{reff5}. According to recent lattice QCD simulations~\cite{karsch07,mueller2}, bulk viscosity $
\xi $ is not negligible near the QCD critical temperature $T_c$. It has been shown
that the bulk and shear viscosity at high temperature $T$ and weak coupling $\alpha_s
$, $\xi\sim  \alpha_s^2 T^3/\ln \alpha_s^{-1}$ and $\eta\sim T^3/(\alpha_s^2 \ln
\alpha_s^{-1})$~\cite{mueller3}. Such a behavior obviously reflects the fact that
near $T_c$ QCD is far from being conformal. But at high $T$, QCD approaches
conformal invariance, which can be indicated by low trace anomaly $(\epsilon-3p)/T^4$
~\cite{karsh09}, where $\epsilon$ and $p$ are energy and pressure density,
respectively. In the quenched lattice QCD, the ratio $\zeta/s$ seems to diverge near
$T_c$~\cite{meyer08}.

To avoid the mathematical difficulties accompanied with the Abel second type non-homogeneous and non-linear differential equations, one used to model the cosmological fluid as an ideal (non-viscous) fluid. No doubt that the viscous treatment of the cosmological background should have many essential consequences~\cite{taw08}.
The thermodynamical ones, for instance, can profoundly modify the dynamics and configurations of the whole cosmological background~\cite{Ma96}. The reason is obvious. The bulk viscosity likely arises due to increasing entropy and can be  expressed as a function of the overall energy density $\rho$~\cite{conseq2}.

In this talk, we assume that the background corresponding to FRW model is filled with ultra-relativistic viscous matter, whose bulk viscosity and equation of state have been deduced from recent heavy-ion collisions experiments and lattice QCD simulations. The relativistic fluid will be described according to the Israel-Stewart (IS) theory.

\section{Model} \label{sec:model}

We take into consideration natural units, i.e., $k=c=1$ and $h=2\pi$ and assume that background geometry of early Universe is filled with bulk viscous cosmic fluid, which can be described by a spatially flat FRW metric. The line element is
\begin{equation}  \label{1}
ds^{2}=dt^{2}-a^{2}(t) \left[dr^{2}+r^{2}\left( d\theta
^{2}+\sin ^{2}\theta d\phi^{2}\right) \right] .
\end{equation}
At vanishing cosmological constant $\Lambda$, Einstein gravitational field equations in a static and flat Universe ($k=0$) read
\begin{equation}
R_{ik}-\frac{1}{2}g_{ik}R=8\pi\; G\; T_{ik}.  \label{ein}
\end{equation}
Inclusion of bulk viscous effects can be generalized through an effective
pressure $\Pi$, which is formally included in the effective thermodynamic pressure $p_{eff}$~\cite{Ma95}. Then in the co-moving frame the energy momentum tensor has the components $T_{0}^{0}=\rho ,T_{1}^{1}=T_{2}^{2}=T_{3}^{3}=-p_{eff}$.
For the line element $a$ given by Eq.~(\ref{1}), we get two independent solutions
\begin{equation}  \label{2}
\left( \frac{\dot{a}}{a}\right) ^{2}=\frac{8\pi}{3}\; G\; \rho ,
\end{equation}
\begin{equation}  \label{3}
\frac{\ddot{a}}{a}=-\frac{4\pi}{3}\; G\; \left( 3p_{eff}+\rho \right) ,
\end{equation}
where one dot denotes differentiation with respect to the comoving time $t$ and $\rho$ is the energy density.
Assuming that the Universe is a closed system, the total energy density of cosmic matter is conserved,  i.e., $T_{i;j}^j=0$.
\begin{equation}  \label{5}
\dot{\rho}+3H\left( p_{eff}+\rho \right) =0,
\end{equation}
Here we introduced the Hubble parameter $H=\dot{a}/a$. 
In the presence of bulk viscous stress $\Pi$, the effective
pressure becomes $p_{eff}=p+\Pi $, where $p$ is the
thermodynamic pressure of the cosmic fluid. Then Eq.~(\ref{5}) can be written as
\begin{equation}  \label{6}
\dot{\rho}+3H\left( p+\rho \right) =-3\Pi H.
\end{equation}
According to the causal theory of relativistic fluid; Israel-Stewart theory~\cite{Is76,IsSt76}, the evolution equation of the bulk viscous pressure reads~\cite{Ma95}
\begin{equation}  \label{8}
\tau \dot{\Pi}+\Pi =-3\xi H-\frac{1}{2}\tau \Pi \left( 3H+\frac{\dot{\tau}}{%
\tau }-\frac{\dot{\xi}}{\xi }-\frac{\dot{T}}{T}\right) ,
\end{equation}
where $T$ is the temperature, $\xi$ denotes the bulk viscosity coefficient and $\tau$ stands for the relaxation time.

Equations of state for $p$ and $T$ can help to have a closed system from Eq.~(\ref{2}), (\ref{6}) and (\ref{8}). $\tau$ and $\xi$ are determined according to phenomenological approaches. For instance, bulk viscosity of QGP at high temperature can be determined approximately from the equilibrium thermodynamical equation of state taken from recent lattice QCD calculations~\cite{karsch07,Cheng:2007jq}.

\begin{equation}\label{13}
p = \omega \rho, \hspace*{1cm} T = \beta \rho^r, \hspace*{1cm} \xi = \alpha \rho + \frac{9}{\omega_0} T_c^4, 
\end{equation}
with $\omega = (\gamma-1)$, $\omega_0 \simeq 0.5-1.5$ GeV, $\beta=0.718$, $\gamma \simeq 1.318$, $r=0.213$ and
\begin{equation}
\alpha = \frac{1}{9\omega_0}  \frac{9\gamma^2-24\gamma+16}{\gamma-1}, 
\end{equation}
From Eq.~(\ref{8}) and (\ref{13}) we obtain an equation describing the cosmological evolution of Hubble parameter $H$
\begin{eqnarray}\label{init}
\ddot H + \frac{3}{2} [1+(1-r) \gamma] H\dot H + \frac{1}{\alpha}\dot H - 
(1+r) \frac{\dot H^2}{H} + \frac{9}{4}(\gamma -2) H^3 + \frac{\gamma}{\alpha} H^2 = 0
\end{eqnarray}
With the transformation $u=\dot{H}$, Eq.~\ref{init} reads
\begin{equation} \label{eq1abel}
u\frac{du}{dH}-(1+r)\frac{u^{2}}{H}+\left\{ \frac{3}{2}[1+(1-r)\gamma]H+\frac{1}{\alpha}\right\} u+\frac{9}{4}(\gamma -2)H^{3}+\frac{3}{2}\frac{\gamma}{\alpha} H^{2}=0
\end{equation}
which is a second kind Abel type first order ordinary differential equation~\cite{abelbook1}.

We take into consideration some numerical approximations~\cite{qgpCosmic1,qgpCosmic2} in order to reduce Eq.~\ref{eq1abel} to an Abel canonical form. We start with the general form of the second kind Abel differential equation
\begin{equation}\label{Abel}
 [y+g(x)]\frac{dy}{dx}=f_2(x)y^2+f_1(x)y+f_0(x), 
\end{equation}
where $g(x)\ne0$. Last equation can be reduced to the Abel canonical form~\cite{abelbook1} by using the following transformations 
\begin{eqnarray}
 \omega &=& [y+g(x)]E \nonumber \\
E &=& \exp\left[-\int f_2(x)dx\right] \nonumber,
\end{eqnarray}
Then the canonical form reads 
\begin{equation}\label{Abel1}
 \omega\frac{d\omega}{dx}=F_1(x)\omega+F_0(x)
\end{equation}
 where,
\begin{eqnarray}
F_1(x) &=& (f_1-2f_2g+g'_x)E \nonumber \\
F_0(x) &=& (f_0-f_1g+f_2g^2)E^2 \nonumber
\end{eqnarray}
Introducing new independent variable $z=\int F_1(x)dx$ in Eq. (\ref{Abel1}) results in
\begin{equation}\label{Abel-cononical}
 \omega\omega'_z-\omega=g(z)
\end{equation}
where $g(z)$ is defined parametrically as~\cite{qgpCosmic1,qgpCosmic2}
\begin{equation}
 g(z)=\frac{F_0}{F_1} \approx {\cal A}\; z + {\cal B}
\end{equation}
where ${\cal A}$ and ${\cal B}$ are constants. For simplicity, we approximate $g(z)$ to ${\cal A}\; z$. 
Then Eq.~\ref{eq1abel} turns to be solved in $H(t)$
\begin{equation} \label{eq:mysolut1}
H(t) = \frac{B}{{\cal E}-A} 
\end{equation}
where denominator is dimensionless, 
\begin{eqnarray}
A=\frac{-3[1+(1-r)\gamma]}{2(1-r)}, \hspace*{1cm} B=\frac{\alpha^{-1}}{r}, \hspace*{1cm} {\cal E}=\exp(-B\;\sigma\;t) \nonumber 
\end{eqnarray}
The dimensionless $\sigma$ is taken as a free parameter. Obviously, it's physical value in the expanding Universe is likely negative. Finally, the scale factor reads,
\begin{equation} \label{eq:mysoluta}
a(t)=a_0\left(\frac{{\cal E}}{{\cal E}-A}\right)^{1/A\,\sigma},
\end{equation}
where $a_0$ is the scale factor determined under the initial conditions. At ${\cal E}>>A$, $a$ remains constant with increasing $t$. The other case that ${\cal E}<<A$ is likely not accessible, no matter which values will be assigned to $\sigma$. Otherwise $a$ slightly increases with increasing the comoving time $t$.

\section{Results and Discussion }\label{sec:discOut}

\begin{figure}
\centering
\includegraphics[width=8cm,angle=-90]{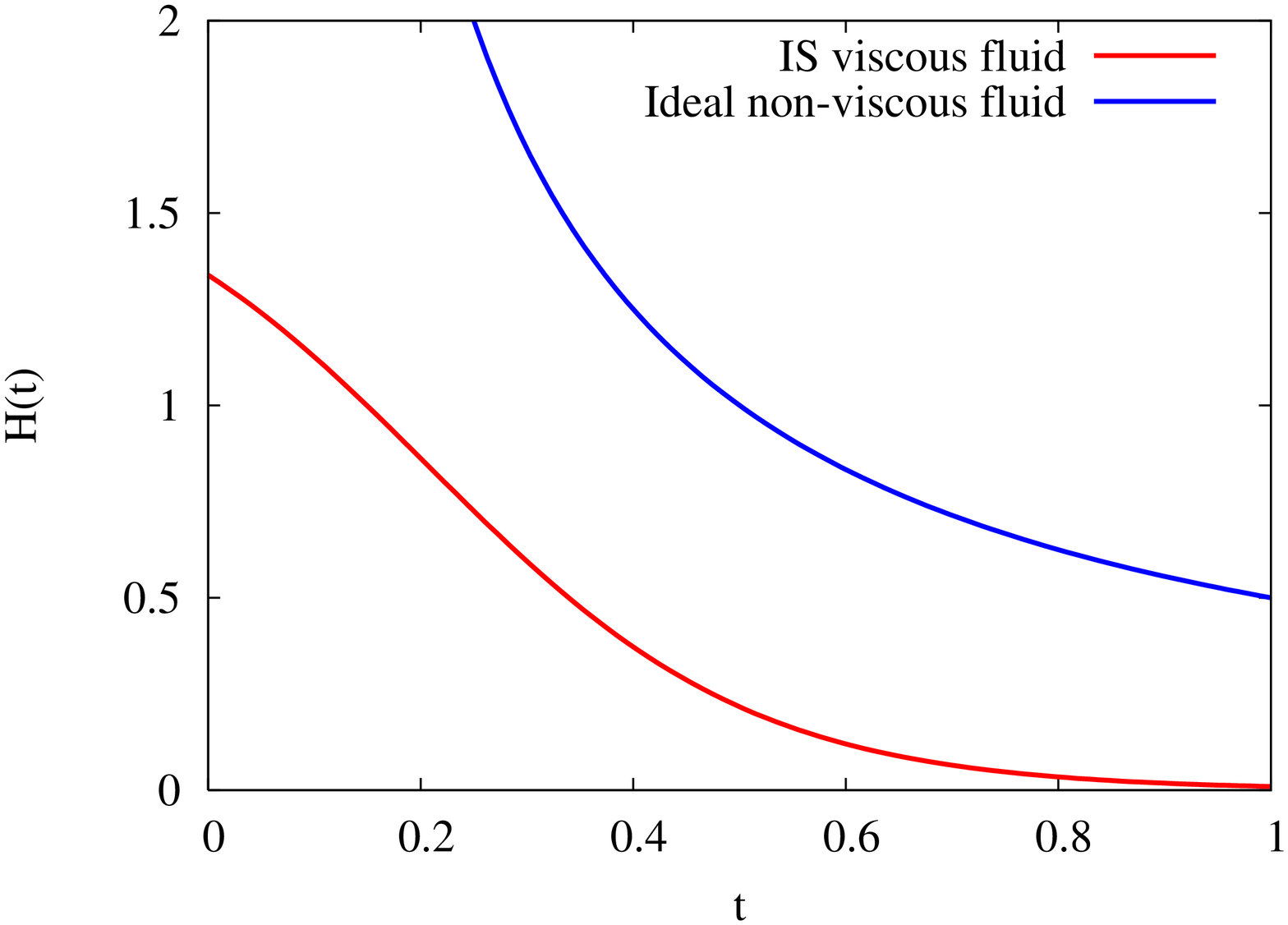}
\includegraphics[width=8cm,angle=-90]{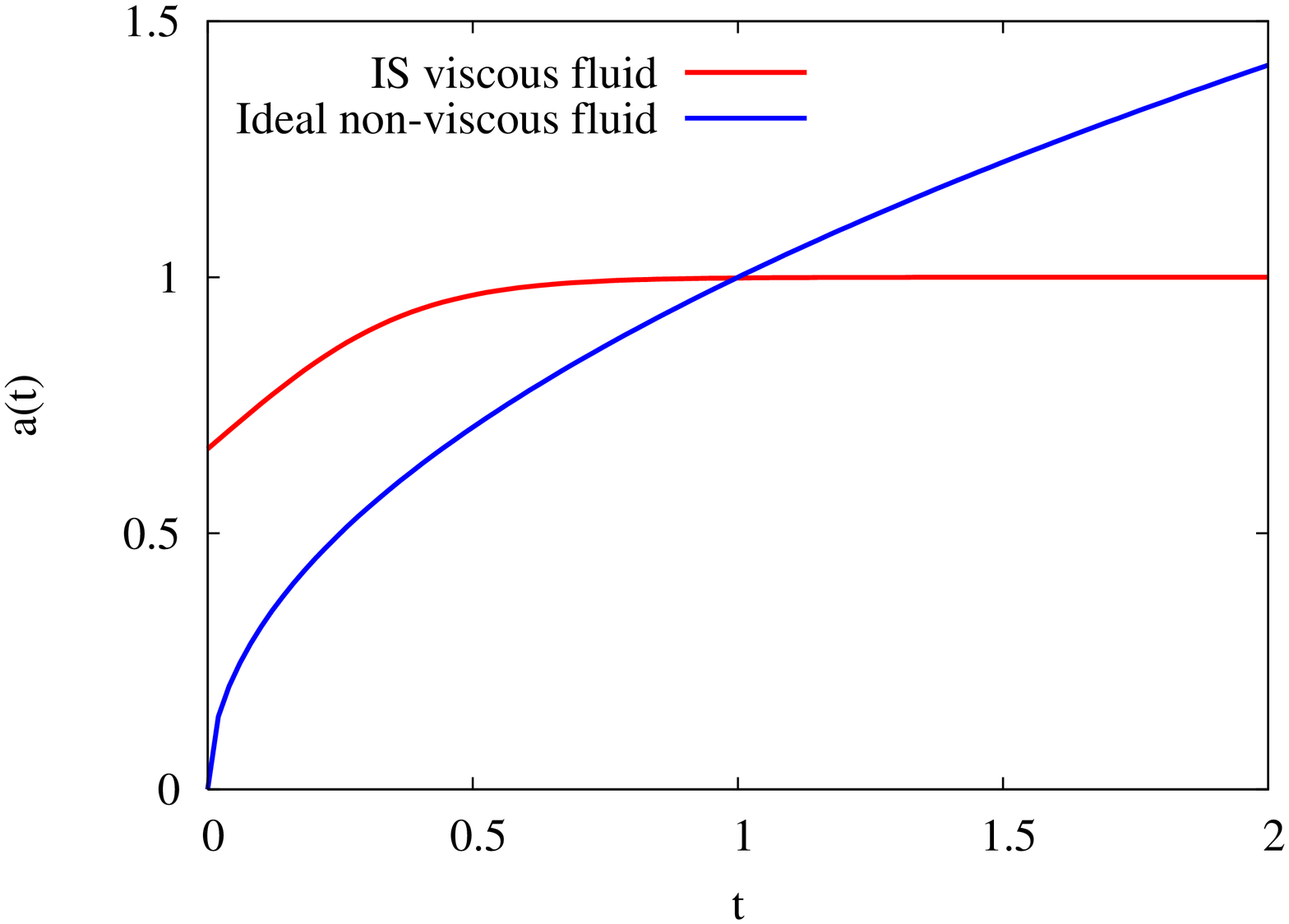}
\caption{Top panel: Hubble parameter of viscous cosmology depicted as function of the comoving time $t$, Eq.~\ref{eq:mysolut1} at $\sigma=-1.0$ and compared with the non-viscous case, $H=1/(2t)$. Bottom panel: the scale factor vs. $t$, Eq.~\ref{eq:mysoluta} at $\sigma=-1$ compared with the ideal behavior, $a(t)=\sqrt{t}$.}
\label{fig1}
\end{figure}

In Fig.~\ref{fig1}, $H(t)$ and $a(t)$ are depicted in dependence on the comoving time $t$. We compare our results, Eq.~\ref{eq:mysolut1} \& \ref{eq:mysoluta}, with $H(t)$ and $a(t)$ resulting in when the background matter is assumed to be filled with an ideal and non-viscous fluid and utilizing equations of state of non-interacting ideal gas
\begin{eqnarray}
H(T) &=& 1/2t \label{htideal} \\
a(t) &=& \sqrt{t} \label{atideal}
\end{eqnarray}
In the case of viscous fluid, it is obvious that $H(t)=\dot a/a$ has an exponential decay, whereas in the non-viscous case $H(t)$ is decreasing according to Eq.~\ref{htideal}. The latter is much slower than the former reflecting the nature of the exponential and linear dependencies. The other difference between the two cases is obvious at small $t$. We notice a divergence or singularity associated with the ideal non-viscous fluid, Eq.~\ref{htideal}. The viscous fluid results in finite $H$ even at vanishing $t$, Eq.~\ref{eq:mysolut1}.   

Also the scale factor $a(t)$ reflects differences in both cases. $a(t)$ in a Universe with an ideal and non-viscous background matter depends on $t$ according to Eq.~\ref{atideal}, which simply implies that $a(t)$ is directly proportional to $t$ and $a(t)$ vanishes at $t=0$ which reflects the singularity of $H(t)=\dot a/a$. Assuming that the background matter is filled out with a viscous fluid results in different $a(t)$-behavior with increasing $t$. At $t=0$, $a(t)$ remains finite. Correspondingly, $H(t)$ remains also finite. In general, the dependence on $t$ is much more complicated than in Eq.~~\ref{atideal}. Here we have an $A\sigma$ root of an exponential function. If ${\cal E}>>A$, $a$ remains constant with increasing $t$.  

As given in section~\ref{sec:model}, to deal with viscous Early Universe, one has to solve the second kind Abel first order ordinary differential equation~\cite{abelbook1}, which is a non-trivial task. In doing this, we have to approximate $g(z)$ in the Abel canonical form, Eq.~\ref{Abel-cononical}. This is the only variable, which we approximated numerically.

In doing this, we assumed that the Universe is flat, $k=0$, and the background matter has a finite viscosity coefficient. The resulting Universe is obviously characterized by a constant scale factor and a vanishing Hubble parameter at large $t$. At small $t$, both $a$ and $H$ remain finite even at vanishing $t$, i.e., no singularity. The validity of our solutions depends on the validity of the equations of states, Eq.~\ref{13}, which we deduced from lattice QCD simulations at temperatures larger than $T_c\approx 0.19~$GeV. Below $T_c$, when the Universe cooled down, not only the degrees of freedom suddenly increase~\cite{Tawfik03} but also  the equations of state turn to be the ones characterizing the hadronic matter. Such a phase transition to hadronic matter would characterize one end of the validity of our solutions. The other end is at very high temperatures, at which the strong coupling $\alpha_s$ entirely vanishes.  \\

\vspace*{1cm}  
  
{\bf \large Acknowledgment}

This work is based on an invited talk given at the 7th international conference on "Modern Problems of Nuclear Physics", 22-25 September 2009 in Tashkent-Uzbekistan. I would like to thank the organizers for their kind hospitality. My special thanks go to Professor Murokha Rasulova for her kind initiative, fruitful discussion and cooperation.


\end{document}